\documentstyle[12pt]{article}
\begin{document}

\flushbottom
\newcommand\ie {{\it i.e. }}
\newcommand\eg {{\it e.g. }}
\newcommand\etc{{\it etc. }}
\newcommand\cf {{\it cf.  }}
\newcommand\viz{{\it viz. }}
\newcommand\grad{\nabla}
\newcommand\noi{\noindent}
\newcommand\seq{\;\;=3D\;\;}
\newcommand\barcaps{\cal}
\newcommand\jump{\vspace*{17pt}}
\newcommand{\R}{{\sf R\hspace*{-0.9ex}\rule{0.15ex}%
       {1.5ex}\hspace*{0.9ex}}}
\newcommand{\N}{{\sf N\hspace*{-1.0ex}\rule{0.15ex}%
       {1.3ex}\hspace*{1.0ex}}}
\newcommand{\Q}{{\sf Q\hspace*{-1.1ex}\rule{0.15ex}%
       {1.5ex}\hspace*{1.1ex}}}
\newcommand{\C}{{\sf C\hspace*{-0.9ex}\rule{0.15ex}%
       {1.3ex}\hspace*{0.9ex}}}
\newcommand\emptypage{~~~ \eject}
\setlength{\baselineskip}{17pt}
\def\be{\begin{eqnarray}}
\def\ee{\end{eqnarray}}
\newenvironment{draftequation}[1]{\be\label{#1}}{\ee}
\newcommand\bbe[1]{\begin{draftequation}{#1}}
\newcommand\eee{\end{draftequation}}
\def\ps{p\hspace{-0.075in}/}
\def\pis{\pi\hspace{-0.075in}/}
\def\half{{\textstyle{1 \over 2}}}
\def\ihalf{{\textstyle{i \over 2}}}
\def\D{{\cal D}}
\newcommand\art[1]{\cite{#1}}
\newcommand\ekv[1]{(\ref{#1})}

\begin{flushright} USITP-96-13\\ September 1996
\end{flushright}

\bigskip
\Large
\begin{center}
\bf{The Mass Spectrum of the 2-dimensional Conformal String}

\bigskip

\normalsize by
\bigskip

P. Saltsidis\footnote{e-mail address:panos@vanosf.physto.se}\\  {\it
ITP\\ University of Stockholm\\ Box 6730, Vanadisv\"agen 9\\ S-113 85
Stockholm\\  SWEDEN}\\
\end{center}
\vspace{2.0cm}
\normalsize
\bigskip
\bigskip

{\bf Abstract:} We present the mass spectrum of the tensionless string 
in 2 dimensions where it has been found that the space time conformal
symmetry  survives quantization. A BRST treatment of the physical
states reveals that the string  collapses into a massless particle, a
result which agrees with the classical treatment.

\eject

\begin{flushleft}
\section{Introduction}
\end{flushleft}

The characteristic energy scale in string theory is set by the level
separation in the  spectrum, which is of order $\sqrt{T}$, where $T$ is
the string tension. The limit $T\rightarrow 0$ corresponds to a  high
energy limit. Understanding this regime of string theory might
contribute to clarify the fundamental structure of quantum strings and
consequently the short-distance structure of space-time. Null strings
were first considered in \cite{sc} and they have been found in
\cite{akul} to correspond to  a collection of particles  moving on a
null geodesic.

The quantum behavior of the tensionless string\footnote{The tensionless
strings currently discussed in the D-brane context have not yet been
described in terms  of an action. It is  therefore unclear  how they
are related to those treated here.} has been investigated
 in several articles and  it  has been found that the Lorentz
invariance is preserved for arbitrary space time dimensions
\cite{liraspsr,gararual,bazh}. However, it was shown  in
\cite{jiulbs,BigT} that the tensionless string is actually
 space-time  conformally invariant. It is this symmetry that replaces
the Weyl invariance in the 
$T\rightarrow 0$ limit, something which also occurs for the massless
particle. Requiring this symmetry to be a fundamental symmetry of the
theory  led to the construction of the conformal string, a string with
manifest space-time conformal symmetry  \cite{ISBE,us}. Using a 
Hamiltonian BRST scheme this model was quantized and  a critical
dimension $d=2$  was found. It is the mass spectrum of this anomaly
free theory that we are  investigating in this article.

The content of the paper is as follows: In Section 2 we present the
classical theory where using the equations of motion we find that the
string collapses to a massless particle. In Section 3 we investigate
the quantum case. There using some general arguments for the physical
vacuum we find that there are two pairs of genuine physical states.
This is examined in the subsequent two subsections where we find that
only a state  that corresponds to a massless particle survives.

\begin{flushleft}
\section{The classical case}
\end{flushleft}

In this section we discuss the $d=2$ bosonic string in some detail
reviewing known results and providing the necessary background. Its
action can be written
\cite{ulbsgt1} 
\be  
S=\int d^{2}\sigma
V^{\alpha}V^{\beta}\gamma_{\alpha\beta}\label{action},
\ee  
where $\gamma_{\alpha\beta}=\partial_{\alpha}
X^{\mu}\partial_{\beta}X_{\mu}$ is the induced metric $\mu =0,1$ is a
space-time index,  $\alpha = 0,1$ is a world sheet index and 
$V^{a}$ is a weight
$w=-\half$ contravariant 2-dimensional vector density. The action is
invariant under
 world-sheet
 diffeomorphisms and space-time conformal transformations
\cite{jiulbs}. Under  diffeomorphisms $X^{\mu}$ transforms as a scalar
field 
\be
\delta_{\epsilon}X^{\mu}= \epsilon\cdot\partial X^{\mu}\nonumber
\ee  and $V^{\alpha}$ as a vector density
\be
\delta_{\epsilon}V^{\alpha}=
-V\cdot\partial\epsilon^{\alpha}+\epsilon\cdot\partial 
V^{\alpha}+\half (\partial\cdot\epsilon )V^{\alpha}.\nonumber
\ee  
There are of course many different gauge choices possible for  the
reparametrization  symmetry. It has been found that the following
transverse gauge is particularly useful
\be  V^{\alpha}=(\upsilon,0)\label{gauge}
\ee  with $\upsilon $ a constant. The transverse gauge corresponds to
the conformal gauge $g_ {\alpha\beta} = e^{\phi}\eta_{\alpha\beta}$ in
the tensile theory. Just as in the tensile
 case there is a residual symmetry that leaves (\ref{gauge}) invariant
\be
\delta\tau &=& f^{\prime} (\sigma )\tau +g(\sigma)\label{res},\\
\delta\sigma &=& f(\sigma),
\ee  
with $f(\sigma)$ and $g(\sigma)$ arbitrary functions. The finite
form looks the same with  different functions $f$ and $g$.

The field equations that follow from the action (\ref{action}) are
\be  
V^{\alpha}\gamma_{\alpha\beta}=0,\qquad
\partial_{\alpha}(V^{\alpha}V^{\beta}
\partial_{\beta}X^{\mu}) =0.\nonumber
\ee  
The first of these equations shows that the induced metric is
degenerate. The second group of the  field equations is most easily
interpreted in  the transverse gauge. They become
\be
\ddot{X}^{\mu}=0,\nonumber\\
\dot{X}^{2}=0,\nonumber\\
\dot{X}\cdot X^{\prime} =0.\nonumber
\ee  These equations give
\be
\frac{d^{2} X^{\mu}}{d\tau^{2}}=0 &\Rightarrow & X^{\mu}(\sigma,\tau)=
F(\sigma)\tau a_{1}^{\mu} +G(\sigma)a^{\mu}_{2} + a^{\mu}_{3}\nonumber\\
\dot{X}^{2}=0 &\Rightarrow &a^{2}_{1}=0\Rightarrow
a^{\mu}_{1}=e^{\mu},\nonumber
\ee  
where $e^{\mu}$ is a null vector pointing in either of the two
light-like directions. The last equation gives
\be
\dot{X}\cdot X^{\prime} =0\Rightarrow e\cdot a_{2} =0 
\Rightarrow a_{2}^{\mu}= e^{\mu}.\nonumber
\ee  
Thus the solution of the equations of motion is 
\be  
X^{\mu}(\sigma,\tau) = [F(\sigma)\tau +G(\sigma)]e^{\mu} +
a^{\mu}_{3}.\nonumber
\ee  
Using the residual symmetry (\ref{res}) we can rewrite this as
\be 
X^{\mu}=\tau^{\prime} e^{\mu} + a^{\mu}_{3},\label{class}
\ee  
which is the equation of a massless particle in two space-time
dimensions, as observed in \cite{us}.

\begin{flushleft}
\section{The quantum case}
\end{flushleft}

Passing to the Hamiltonian formulation we  find, \cite{us}, that the
Hamiltonian is a linear combination of the two constraints
\be
\phi^{-1}(\sigma) =P^{\mu}P_{\mu}(\sigma)=0,\qquad
\phi^{L}(\sigma)=P^{\mu}X_{\mu}^{\prime} =0,\nonumber
\ee
 which is expected since the theory is reparametrization invariant. In
Fourier modes the  constraints read
\be
\phi^{-1}_{m}&=&\half \sum_{k=-\infty}^{+\infty}p_{k}\cdot p_{m-k}=
0,\label{ccon1}\\
\phi^{L}_{m}&=&-\frac{i}{2}\sum_{k=-\infty}^{+\infty}[k x_{k}\cdot
p_{m-k}+
 kp_{m-k}\cdot x_{k}]= 0\label{ccon2}.
\ee  and they satisfy the following algebra
\be
\left[ {\phi}^{-1}_{m},{\phi}^{L}_{n}\right]   &=&
                (m-n){\phi}^{-1}_{m+n},\label{excont}\\
  \left[ {\phi}^{L}_{m},{\phi}^{L}_{n}\right] &=&(m-n)
    {\phi}^{L}_{m+n},
   \label{d12} 
\ee  
where the basic non zero commutators are
\be
\left [x^{\mu}_m,p^{\nu}_n\right ] =i\delta_{m+n}\eta^{\mu\nu}.\nonumber
\ee  
The BRST quantization requires the introduction   of new
operators, the  Faddeev-Popov ghosts. To every constraint
$\phi^{A}_{m}$, $A\in\{-1,L\}$ one introduces a ghost pair
 $c^{A}_{m}$, $b^{A}_{m}$ that is fermionic. The ghosts satisfy the
fundamental anticommutation relations
\be
\left \{c^{A}_m,b^{B}_n\right \} =\delta_{m+n}\delta^{AB}.\nonumber
\ee  The generator of BRST transformations, the BRST charge is given by
\cite{FRAD}
\be
 Q = \sum_{k} (\phi_{-k}^{-1}c_{k}^{-1}+
\phi^{L}_{-k}c_{k}^{L}) -
\sum_{k,l} [(k-l)c_{-k}^{-1}c_{-l}^{L} b_{k+l}^{-1}+\half
(k-l)c_{-k}^{L}c_{-l}^{L}b_{k+l}^{L}].\label{Q}
\ee  
This BRST operator was found to be nilpotent in \cite{liraspsr}.
 The physical states satisfy the  condition
\be  Q|phys\rangle =0.\label{con}
\ee The nilpotency of $Q$ implies that any state of the form $
Q|\rangle$ is a physical state. These states are called exact states
and due to (\ref{con}) they will decouple from all physical states
i.e., have inner product equal to zero. The physics is therefore
contained in equivalence classes of physical states (BRST  cohomology).
In the analysis of the BRST condition (\ref{con})  it is convenient to
introduce the ghost number operator
\be  
N_{gh}=\half \sum_{k}\left [c^{-1}_{-k}b^{-1}_{k}
+c^{L}_{-k}b^{L}_{k} -b^{-1}_{-k}c^{-1}_{k}-b^{L}_{-k}c^{L}_{k}\right
].\label{ghn}
\ee  
It satisfies
\be 
\left [N_{gh},c^{A}_{k}\right ] = c^{A}_{k},\qquad 
\left [N_{gh},b^{A}_{k}\right ] = -b^{A}_{k}\nonumber
\ee
 and also 
\be
\left [N_{gh},Q\right ] = Q.\nonumber
\ee  
We can, therefore, classify physical states according to their
ghost number
\be  
N_{gh} |phys,n\rangle =n|phys,n\rangle .\nonumber
\ee  
Since $N_{gh}$ is an antihermitian operator with real eigenvalues,
the physical states should satisfy
\be
\langle phys, m|phys,n\rangle = C_{n}\delta_{m+n}.\nonumber
\ee

In \cite{MarnBRST} a method of solving the physical state condition
(\ref{con})
 was proposed. It is argued that it is sufficient to consider states of
the form
\be 
|matter\rangle |ghost\rangle .\nonumber
\ee 
A fundamental prescription of this method is the requirement that
all inner products have to be finite. As a consequence the genuine
physical states will appear in pairs: to every  physical state or the
form $|phys \rangle =|M\rangle |G\rangle $ there is a dual physical
{\it bra} state of the form $\langle {phys}^{\prime}|=\langle
-G|\langle M^{\prime}|$.
 We choose the normalization of the ghost states to be $\langle
-G|G\rangle =1$. The ghost state space is built from the vacuum states
$|0\rangle_{c}$ and
$|0\rangle_{b}$,
 which satisfy
\be  
c^{A}_{k}|0\rangle_{c} & = &  0,\nonumber\\ b^{A}_{k}|0\rangle_{b}
& = & 0,\qquad \forall k,\nonumber
\ee
\be
\mbox{}_{c}\langle 0|0\rangle_{c}=\mbox{}_{b}\langle
0|0\rangle_{b}=0,\qquad 
\mbox{}_{c}\langle 0|0\rangle_{b}=1.\label{sat}
\ee  
The two ghost vacua are related by the equation 
\be 
|0\rangle_{c}
=N\prod_{k=-\infty}^{+\infty}c^{-1}_{k}\prod_{l=-\infty}^{+\infty}c^{L}_{l}
|0\rangle_{b},
\ee 
where  $N$ is a constant chosen to satisfy (\ref{sat}). Applying
the operator (\ref{ghn}) to the states
$|0\rangle_{c}$ and
$|0\rangle_{b}$
 we find that their ghost number, $C$ and $B$, is $+\infty$ and
$-\infty$ respectively.  An arbitrary ghost state will have the form
\be  
|B +k\rangle^{a_1\ldots a_k}  =  c^{a_1}\ldots
c^{a_k}|0\rangle_{b}\nonumber
\ee  
or equivalently
\be  
|C -k\rangle^{a_1\ldots a_k} & = & b^{a_1}\ldots
b^{a_k}|0\rangle_{c},\nonumber
\ee  
where $a_k = (A,k)$, $A\in \{-1,L \}$ and $k$ any integer. Assume
now  the {\it ket} state
$|G^{-1}_{s}\rangle\equiv c^{-1}_s |0\rangle_{b}$ which satisfies the
condition 
$c^{-1}_s|G^{-1}_{s}\rangle =0$. Non-zero inner products can be
constructed only with {\it bra} states of the form $\mbox{}_{c}\langle
0|b^{-1}_{-s}$ since\footnote{States of the form $\mbox{}_{c}\langle 0|
b^{a_1}\ldots b^{a_k}$ with more than one $b^{a}$ operator  give zero
inner products with $|G^{-1}_{s}\rangle$ since we can always move one
of the $b^{a}$'s to the right of $c^{-1}_{s}$.}
\be
\mbox{}_{c}\langle 0|b^{A}_{m}|G^{-1}_{s}\rangle
=\delta_{A,-1}\delta_{m+s}.
\ee  
The state $\langle -G^{-1}_{-s}|\equiv \mbox{}_{c}\langle
0|b^{-1}_{-s}$ satisfies the relation
$\langle -G^{-1}_{-s}|b^{-1}_{-s}=0$. In the same way we can prove for
a general ghost state $|G\rangle$
\be  c^{A,k}|G\rangle =0\Rightarrow \langle
-G|b^{A,-k}=0.\label{ghostrel}
\ee
 We find it useful to define also the following states 
\be 
c^{-1}_{k}|G^{-1}_{M}\rangle & = & c^{L}_{k}|G^{L}_{M}\rangle =
0,\nonumber\\ b^{-1}_{k}|-G^{-1}_{M}\rangle &= &
b^{L}_{k}|-G^{L}_{M}\rangle = 0.\label{def}
\ee

According to arguments presented in \cite{BigT}  the  vacuum suitable
for  tensionless strings is not   one annihilated by the positive modes
of the operators but  one annihilated  by the momenta\footnote{cf. the
vacuum for a particle.}
\be
  p^{\mu}_{m}|0\rangle_{p} =0 \quad \forall m.\label{vac}
\ee  
Following the prescription of \cite{MarnBRST}, we will take  the
{\em ket } states to be built from our vacuum of choice,
$|0\rangle_{p}$, and the {\em bra } states to be built from
$\mbox{}_{x}\langle 0|$ satisfying
$\mbox{}_{x}\langle 0|0\rangle_{p}=1$. Asymmetric  states for the
coordinates and the above restriction may be constructed in  different
ways. We will make use of the so called wave sector, with the {\it bra}
states defined by
\be
\langle x|\qquad \mbox{or}\qquad
\mbox{}_{x}\langle 0|Y(p)\label{wf}
\ee  
and the {\it ket} states defined by
\be
\Phi(x)|0\rangle_p\qquad\mbox{or}\qquad \qquad |p\rangle
\ee  
respectively. In this relations $|x\rangle$ and $|p\rangle $
represent the collective $|\ldots,x_{n},\ldots\rangle$ and
$|\ldots,p_{n},\ldots\rangle$ respectively.

For the vacuum (\ref{vac}) and from the requirement that the BRST
charge (\ref{Q}) should annihilate the vacuum, we obtain further
requirements on the ghost part of the vacuum. Defining $|phys\rangle
_{0}=|0\rangle_{p}|G\rangle$ we have
\be  Q|phys\rangle _{0} =0\Rightarrow  \left [Q,p^{\mu}_{n}\right
]|phys\rangle _{0}=0,\nonumber
\ee  
while
\be
\left [Q,p^{\mu}_{n}\right ]|phys\rangle _{0}
 = \sum_{k}(-n)c^{L}_{-k}p^{\mu}_{k+n}|0\rangle_{p}|G\rangle=0,\nonumber
\ee  
with the last relation  trivially satisfied. In the same way,
requiring the {\it bra} vacuum $\mbox{}_{0}\langle phys|\equiv
\mbox{}_{x}\langle 0|\langle -G|$ to be a  physical state  restricts 
the ghost part of the state
\be
\mbox{}_{0}\langle phys| \left [Q,x^{\mu}_{n}\right ]= 0 &\Rightarrow &
\mbox{}_{x}\langle 0|\langle -G|\sum_{k}\left
[c^{-1}_{-k}(-i)p^{\mu}_{k+n}+ c_{-k}^{L}(-k-n)x^{\mu}_{k+n}\right
]=0\nonumber\\
 &\Rightarrow &\langle -G|c^{-1}_{k}=0\Rightarrow b^{-1}_{k}|G\rangle
=0,
 \qquad \forall k\label{gvac}.
\ee  
This means that any physical state $|phys\rangle
=|M\rangle|G\rangle$ should have a ghost part that satisfies the
relations (\ref{gvac}). We  get further conditions on the physical
states by requiring the consistency relations
\be
\langle M^{\prime}|\langle -G|\left [ Q, c^{-1}_{m}\right ]= 0,
\qquad \left [Q,b^{-1}_{m}\right ]|M\rangle|G\rangle =0.\nonumber
\ee  
The first one is trivially satisfied. The second  restricts  the 
matter part. It requires
\be 
[\phi^{-1}_{m}-\sum_{l}c^{L}_{-l}b^{-1}_{m+l} ]|M
\rangle |G\rangle =0\Rightarrow\nonumber\\
\phi^{-1}_{m}|M\rangle =0,\qquad\forall m.\nonumber
\ee
 We note that $\left[Q,b^{-1}_{m}\right ]\equiv \tilde{\phi}^{-1}_{m}$
are the extended constraints defined in \cite{us}, i.e. the BRST
invariant extensions of  the original constraints and they satisfy the
same algebra. They are very  useful in the calculation of the BRST
anomaly.

So far, using general arguments on the physical vacuum we found that 
the physical states  have to be of the form
\be  
|phys\rangle = |M\rangle |G\rangle,\qquad \langle phys^{\prime}|=
\langle M^{\prime}|\langle -G|\nonumber
\ee
 with the ghost states constrained  by the conditions (\ref{gvac}).
However, there is still a freedom in the ghost part  since it can have
the form $\prod_k b^{L}_{k} |-G^{-1}_{M}\rangle$  where $k$ belongs to
any set of non repetitive integers. This suggests that there is an
infinite number of ghost sectors. That this is not true can be seen as
follows: Assume that the ghost part of the physical state satisfies the
condition
\be  
b^{L}_{s}|G\rangle =0\nonumber
\ee  
for a specific $s\neq 0$. Thus, we also have (\ref{ghostrel})
\be
\langle -G|c^{L}_{s}=0.\nonumber
\ee
 The consistency conditions also require  the following relations to
hold
\be
 &&\langle M^{\prime}|\langle -G|\left [ Q, c^{L}_{s}\right ]=
0\Rightarrow
\langle -G|\sum_{k} (2k+s)c^{L}_{-k}c^{L}_{k+s} =0\nonumber\\
&\Rightarrow &\langle -G|c^{L}_{k}=0
\Rightarrow b^{L}_{k}|G\rangle =0,\qquad \forall k\neq 0\nonumber
\ee  
and
\be
\left [Q,b^{L}_{m}\right ]|phys\rangle =0&\Rightarrow &
[\phi^{L}_{m}-2mc^{L}_{m}b_{0}^{L}]|M\rangle |G\rangle\nonumber\\
 \phi^{L}_{m}|M\rangle =0, && b^{L}_{m}|G\rangle =0,\qquad\forall
m.\label{con12}
\ee  
As a result, by requiring the physical state to be annihilated by
just one operator 
 $b^{L}_{s}$, consistency conditions force the relation (\ref{con12})
for all $m$.
 In the same way we can prove that by requiring  the ghost part of a
physical state to satisfy the condition $b^{L}_{s}|G\rangle =0$ just
for a specific $s$,  the physical state is forced to satisfy the
relations
\be
\langle M^{\prime}|\phi^{L}_m=0,\qquad c^{L}_{m}|G\rangle
=0,\qquad\forall m.\nonumber
\ee  
Therefore, gathering all the previous results together we find
that there are two pairs of  genuine physical states. One pair has
infinite ghost number and is given by (\ref{def})
\be  
|phys,+\infty\rangle &=& |M\rangle |-G^{-1}_{M},-G^{L}_{M}\rangle
,\nonumber \\
\langle phys^{\prime},-\infty| & = & \langle M^{\prime}|\langle
G^{-1}_{M}, G^{L}_{M}|,\nonumber
\ee  
with  the matter part satisfying
\be
\phi^{-1}_m |M\rangle =\phi^{L}_m |M\rangle =0,\qquad \forall
m.\label{c1}\
\ee  
The other pair has ghost number zero and is given by
\be  
|phys,0\rangle &=& |M\rangle |-G^{-1}_{M},G^{L}_{M}\rangle
,\nonumber\\
\langle phys^{\prime},0| & = & \langle M^{\prime}|\langle G^{-1}_{M},
-G^{L}_{M}|,\nonumber
\ee  
with  the matter part satisfying
\be
\phi^{-1}_m |M\rangle  =0,&& \label{cc1}\\
\langle M^{\prime}|\phi^{L}_{m} =0,&&\forall m.\label{c2}
\ee  
We will  now  investigate these two cases in detail.

\begin{flushleft}
\subsection{Case I}
\end{flushleft}

The conditions on the physical states in this case are the same as the
ones that appear in  Dirac quantization of the model. In the wave
function sector (\ref{wf}), the solution may be written as
\be  
|M\rangle &=& \Phi (x)|0\rangle_p,\nonumber\\
\langle M^{\prime}|& =& \langle x|\nonumber
\ee  
and the inner product gives
\be
\langle phys^{\prime},+\infty|phys, -\infty\rangle =\Phi (x).\nonumber
\ee  
Equation (\ref{c1}) then implies the equations
\be 
\sum_{k=-\infty}^{+\infty}\frac{\partial}{\partial
x^{\mu}_{k}}
\frac{\partial}{\partial {x_{-k-m}}_{\mu}}\Phi (x) &=&0,\label{g1}\\
\sum^{+\infty}_{k=-\infty}k x^{\mu}_k
\frac{\partial}{\partial {x_{k-m}}^{\mu}}
\Phi (x) &=&0\label{g2}.
\ee

Before we continue to   the solution for general $m$  let us take a
closer look at  the conditions (\ref{c1}) for $m=0$.
\be  
&&\phi^{-1}_{0}|M\rangle =\half\sum_{k}p_{k}\cdot p_{-k}|M\rangle =
\half p_{0}^{2}|M\rangle +\sum_{k>0}p_{k}\cdot p_{-k}|M\rangle =
0\nonumber\\ 
&\Rightarrow & -p_{0}^{2}|M\rangle = 2\sum_{k>0}p_{k}\cdot
p_{-k}|M\rangle
\Rightarrow M^{2} |M\rangle = 2\sum_{k>0}M^{2}_{k}|M\rangle
,\label{mass}
\ee  
where we have used the definition of  the mass operator
$M^{2}=-p_{0}^{2}$ and defined $p_{k}\cdot p_{-k}
\equiv M^{2}_{k}$. We also have
\be
\phi^{L}_{0} =-i\sum_{k>0}^{+\infty}k(x_k\cdot p_{-k}-x_{-k}\cdot
p_{k})\equiv 
\sum_{k>0}^{+\infty}k S_k.\nonumber
\ee  
It turns out that the commutators $ [M^{2},\phi^{L}_{0}]$,
$ [M^{2},S_k ]$, $ [M^{2},M^{2}_k ]$,
$ [\phi^{L}_{0},S_k ]$, $ [\phi^{L}_{0},M^{2}_k ]$,
$ [M^{2}_k,S_l ]$, $ [S_k,S_l ]$,
$ [M^{2}_k,M^{2}_l ]$ vanish. The
 problem of finding mass eigenstates can, therefore, be reduced to that
of  simultaneously diagonalizing $S_k$ and $M^{2}_k$
\be  
S_k \Phi_k = s_k \Phi_k\mbox{ and }{M}^{2}_{k}\Phi_k = m^{2}_k
\Phi_k.\nonumber
\ee  
The wave function of a generic state will be written as a product
of functions, a factor for each $k\leq 0$
\be
\Phi (x) = \prod_{k=0}^{+\infty}\Phi_k (x_k,x_{-k})\label{tot}.
\ee  
The zero-modes  $p^{\mu}_{0}$, $x^{\mu}_{0}$ are not contained in
$\phi^{L}_{0}$ and 
$M^{2}$. To them corresponds a plane wave describing the motion of the
string's center of mass
\be
\Phi _{k=0}=e^{il_0\cdot x_0}.\nonumber
\ee  
We  now look for the other functions ($n\neq 0$). They have to
satisfy the  equations
\be  
-\frac{\partial}{\partial x^{\mu}_{k}}
\frac{\partial}{\partial {x_{-k}}_{\mu}}\Phi_{k} (x_k,x_{-k})  &=&
m^{2}_k \Phi_{k} (x_k,x_{-k}),\label{1}\\
 \left ( x^{\mu}_{-k} \frac{\partial}{\partial {x_{-k}}^{\mu}} -
x^{\mu}_k
\frac{\partial}{\partial {x_{k}}^{\mu}}\right )
\Phi_{k} (x_k,x_{-k}) &=& s_k\Phi_{k} (x_k,x_{-k}).\label{2}
\ee  
In order to solve the equations (\ref{1}) and (\ref{2}) we notice
that the operator
$X(\sigma )$ is Hermitian. This means that the modes
$x_k$ and $x_{-k}$ are related by the equation ${x_k}^{\dagger}=
x_{-k}$ and correspondingly,  they are represented in the wave sector
by
\be
x_k \to x_k,\qquad x_{-k}\to x_k^{\ast},\qquad k>0\nonumber.
\ee
where $x_k^{\ast}$ is the complex conjugate of $x_k$.
For a real wave function the equation (\ref{2})
is trivially satisfied. Hence we
 can
distinguish between two cases. In the first  we take the wave  function
to be  real  i.e. we assume that $x_k=x_{-k}$. In the second,
which will be investigated later, we allow $x_{-k}\neq x_k$. 
 Applying (\ref{1}) to the first case we get
\be
&&\frac{\partial}{\partial x^{\mu}_{k}}
\frac{\partial}{\partial {x_{k}}_{\mu}}\Phi_{k} (x_k)  =
-m^{2}_k \Phi_{k} (x_k)\nonumber\\
&\Rightarrow &
\Phi_{k}(x_k) = e^{il_{k}\cdot x_{k}},\nonumber
\ee  
with the condition $l_{k}\cdot l_{k}= m^{2}_{k}$. The total wave
function (\ref{tot})  should  then have the form
\be
\Phi (x) = \exp\left (i\sum_{k=0}^{s}l_{k}\cdot x_{k}\right
),\nonumber
\ee  
where we have taken $s$ to be a very large positive integer. This
wave function has to satisfy the equations  (\ref{g1}) and (\ref{g2})
for any integer $m$. Those equations are trivially satisfied for
$m>2s$. For $m=2s$ (\ref{g1}) gives
\be
\frac{\partial}{\partial x^{\mu}_{s}}
\frac{\partial}{\partial {x_{s}}_{\mu}}
\exp\left (i\sum_{k=0}^{s}l_{k}\cdot x_{k}\right )=0\Rightarrow
l^{2}_{s}=0.\nonumber
\ee  
For $m=2s-1$ we have
\be  
&&\left (\frac{\partial}{\partial x^{\mu}_{s}}
\frac{\partial}{\partial {x_{s-1}}_{\mu}}+
\frac{\partial}{\partial x^{\mu}_{s-1}}
\frac{\partial}{\partial {x_{s}}_{\mu}}\right )
\exp\left (i\sum_{k=0}^{s}l_{k}\cdot x_{k}\right )=0\nonumber\\
 &\Rightarrow& l_{s}\cdot l_{s-1}=0.\nonumber
\ee  
But in two space-time dimensions 
\be
\left.
\begin{array}{rcl}  l^{2}_{s}&=&0\\  l_{s}\cdot l_{s-1}&=&0
\end{array}
\right\}\Rightarrow l^{2}_{s-1}=0,\nonumber
\ee  
both pointing in the same light-like direction. In the same way we
find that $l_{i}\cdot l_{j} = 0$,
$\forall i,j\in[t,s]$,where $t$ is any integer lesser than $s$. Then
for $m=2s-t-1$
\be
\langle x|\phi^{-1}_{2s-t-1}|M\rangle =0&\Rightarrow & l_{s}\cdot
l_{s-t-1} + l_{s-1}\cdot l_{s-t}+\ldots +l_{s-t-1}\cdot l_{s}=0
\nonumber\\ 
&\Rightarrow & l_{s}\cdot l_{s-1}=0 \Rightarrow
l^{2}_{s-t-1}=0,\nonumber
\ee  
so $l_{n}\cdot l_{m}=0,\forall n,m<s$ and accordingly the states
are {\it{massless}} since (\ref{mass}) gives
\be  
M^{2}\Phi (x) = 2\sum_{k=1}^{s}l_{k}\cdot l_{k}\Phi (x) =
0.\nonumber
\ee

A physical state should also satisfy the relations (\ref{g2}). For
$m=2s$
\be  
&&s x^{\mu}_s \frac{\partial}{\partial {x_{s}}^{\mu}}
\exp\left (i\sum_{k=0}^{s}l_{k}\cdot x_{k}\right )=0
\Rightarrow sx_{s}\cdot l_{s}=0\nonumber\\  
&\Rightarrow &
x^{2}=0,\qquad x_{s}\cdot l_k=0,\qquad \forall k<s.\nonumber
\ee  
In the same way we can prove 
\be  
l_k\cdot x_{m}= 0, \qquad\forall m\neq 0.\nonumber
\ee  
Thus, a physical state can only have the form
\be
\Phi (x) = C e^{il_0\cdot x_0},\nonumber
\ee  
where $C$ is a constant. This is the wave function of a {\it
massless particle}.

We now investigate the second case where we take $x_k\neq x_{-k}$.  To
construct the eigenstates of $S_k$, we start from a general Lorentz
invariant wave function of the form
\be
\Phi(x_k,x_{-k})\equiv Z_k (x_k,x_{-k} )=F_k(l^{k}\cdot
x_k)G_k(l^{k}_q\cdot x_{-k})H_k(x^2_{k})I_k(x^2_{-k})\Psi_k
(\chi_k),\nonumber
\ee
where $\chi_k\equiv x_k\cdot x_{-k}$. This function should satisfy eq.
(\ref{2})

\be  
&&S_k Z_k (x_k,x_{-k} )=s_k Z_k (x_k,x_{-k} )\nonumber\\
&\Rightarrow &\left ( x^{\mu}_{-k}
\frac{\partial}{\partial {x_{-k}}^{\mu}} - x^{\mu}_k
\frac{\partial}{\partial {x_{k}}^{\mu}}\right ) F_k G_k
H_k I_k \Psi_k   = s_k
 F_k G_k
H_k I_k \Psi_k\nonumber\\
&\Rightarrow &
-\frac{(l^{k}\cdot
x_k)}{F_k}\frac{dF_k}{d(l^{k}\cdot x_k)}
+\frac{(l^{k}_{q}\cdot
x_{-k})}{G_k}\frac{dG_k}{d(l^{k}_q\cdot
x_{-k})}\nonumber\\
&&-\frac{2(x^2_{k})}{H_k}\frac{dH_k}{d(x^2_{k})}
+\frac{2(x^2_{-k})}{I_k}\frac{dI_k}{d(x^2_{-k})}=s_k
\nonumber
\ee
Thus the eigenfunctions of  
 $S_k$ should have the form
\be  
Z_k  = (l^{k}_1\cdot x_k )\ldots
(l^{k}_{q_k}\cdot x_k ) (l_{{q_k}+1}^k\cdot x_{-k} )\ldots
(l_{{q_k}+{r_k}}^k\cdot x_{-k} )(x^2_{k})^{n_k}(x^2_{-k})^{g_k}\Psi_k
(\chi_k).\nonumber
\ee 
The corresponding eigenvalue is $s_k =   r_k - q_k -2 n_k + 2 g_k$.
To simplify the notation we define 
$A_k \equiv(l^{k}_1\cdot x_k )\ldots (l^{k}_{q_k}\cdot x_k )$,
$B_k \equiv (l_{{q_k}+1}^k\cdot x_{-k} )\ldots (l_{{q_k}+{r_k}}^k\cdot
x_{-k} )$,
$\hat{A}^{s}_k \equiv(l^{k}_1\cdot x_k )\ldots(l_{s-1}^k\cdot x_k
)(l_{s+1}^k\cdot x_k )
\ldots (l^{k}_{q_k}\cdot x_k )$ and 
$\hat{B}^t_k \equiv (l_{{q_k}+1}^k\cdot x_{-k} )\ldots
(l_{{q_k}+t-1}^k\cdot x_{-k} ) (l_{{q_k}+t+1}^k\cdot x_{-k} )\ldots
(l_{{q_k}+{r_k}}^k\cdot x_{-k} )$.
 Demanding  the function $ Z_k$ to be an eigenfunction
of  the operator (\ref{1}) we find that $n_k = g_k = 0$. We also have
\be  
&&\frac{\partial}{\partial x^{\mu}_{k}}
\frac{\partial}{\partial {x_{-k}}_{\mu}}Z_k (x_k,x_{-k}
)=-m^{2}_{k}Z_k(x_k,x_{-k} )
\nonumber\\  
&\Rightarrow &
\sum_{s=1}^{q_k}\sum_{t=1}^{r_k}l^{k}_{s}\cdot l^{k}_{q_k
+t}\hat{A}^{s}_k 
\hat{B}^{t}_k \Psi_k (\chi_k)
 +A_k B_k
\sum_{s=1}^{q_k}\frac{\partial\Psi_k}{\partial\chi_k}\nonumber\\ 
&&+A_k B_k
\sum_{s=1}^{r_k}\frac{\partial\Psi_k}{\partial\chi_k} + A_k B_k
\frac{\partial}{\partial x^{\mu}_k }
\left [x^{\mu}_k \frac{\partial\Psi_k}{\partial \chi_k}\right ]  =
-m^{2}_{k}A_k B_k
\Psi_k (\chi_k ).\nonumber
\ee  
Thus $l^{k}_{s}\cdot l^{k}_{q_k +t} =0$, $\forall s,t$ and
$\Psi_k$ has to satisfy the equation
\be
\chi_k \Psi^{\prime\prime}_k + (q_k +r_k +d)\Psi^{\prime}_k
+m^{2}_{k}\Psi_k =0,\nonumber
\ee  
where $d$ is the space-time dimension. The solution of this
equation has the form 
\cite{gararual,Bessel}
\be
\Psi_k^{m} (\chi_k) =N_m \chi_k^{-(1+q_k +r_k )/2}J_{1+q_k +r_k }(2m_k
\sqrt{\chi_k}),\nonumber
\ee  
where $J_{1+q_k +r_k }$ is a Bessel  function of rank $1+q_k
+r_k$.  Thus an eigenfunction that
 simultaneously diagonalizes (\ref{1}) and (\ref{2}) has the form
\be  
Z_k^{m} (x_k, x_{-k})=N(l^{k}_1\cdot x_k)\ldots (l^{k}_{q_k +r_k
}\cdot x_{-k})
\chi_{k}^{-(1+q_{k}+r_{k})/2} J_{1+q_{k}+r_{k}}(2m_k
\chi_{k}^{1/2}).\nonumber
\ee  
Since $\hat{m}^{2}_k = p_k\cdot p_{-k}$ is a Hermitian operator
its eigenvalues
$m^{2}_{k}$ should be real  numbers. If $m^{2}_{k}< 0\Rightarrow m_k$
is purely imaginary. Take $m_k =i\mu_k $ with $\mu_k$ real.
 Then
\be
\Psi_k^m (\chi_k) = N_m {(i \mu_k)}^{1+q_k + r_k} \sum^{+\infty}_{n=0}
\frac{(\mu_k \chi_k^{1/2})^{2n}}{n!\Gamma (q_k + r_k +n+2)}.\nonumber
\ee  
According to the latter, for the tachyonic solutions every term in
the same factor
 is positive and hence  the corresponding wave functions cannot be
normalizable. This  means that $m^{2}_k$ should be real and
positive and so  the mass spectrum is continuous ranging from zero to
infinity.

Special consideration is needed for the case $m^{2}_{k}=0$. The
corresponding equation is then
\be
\chi_k \Psi^{\prime\prime}_k + (q_k +r_k +2)\Psi^{\prime}_k  =0
\Rightarrow Z_k^0 (\chi_k) = D_k {\chi_k}^{-(1+q_k +r_k )}
+C_k,\label{massl}
\ee  
where $D_k$ and $C_k$ are constants.

We will assume again the wave function to depend on a finite number of
modes i.e
\be
\Phi (x) =\prod_{k=0}^{K} Z_{k}(x_k,x_{-k}),\label{state}
\ee  
where $K$ is a large but finite integer. Since (\ref{state}) is
the wave function of a physical state it should also satisfy the
equations  (\ref{g1}) for any
$m$. The first non trivially satisfied constraint is again
$\phi^{-1}_{2K}$. This gives
\be  
&&\frac{\partial}{\partial x^{\mu}_{-K}}
\frac{\partial}{\partial {x_{-K}}_{\mu}}
\prod^{K}_{k=1}A_k B_k e^{il_0\cdot x_0}\Psi_k (\chi_k )=0\nonumber\\ 
&\Rightarrow &
\sum_{s=1}^{r_K}\sum_{t\neq s}^{r_K}l^{K}_{q_K +s}\cdot l^{K}_{q_K
+t}\hat{B}^{st}_K \Psi_K (\chi_K)
 +2 \sum_{s=1}^{r_K}l^{K}_{q_K +s}\cdot
x_{K}\hat{B}^{s}_{K}\Psi^{\prime}(\chi_K)\nonumber\\  
&&+B_K x^{2}_K
\Psi^{\prime\prime}_K (\chi_K) =0.\label{tr}
\ee  
This equation holds if all of its terms are equal to zero. In
particular it requires that 
$l^{K}_{q_K +s}\cdot l^{K}_{q_K +t}=0$,
$\forall s\neq t \in [1,r_K]$ and 
$l^{K}_{q_K +s}\cdot x_{K} =0$, $\forall s\in [1,r_K]$.
Working in the same way for all $m$, we find that the constraints 
$\phi^{-1}_m$ imply the conditions
 $l^{k}_{s}\cdot l^{n}_{t}=
l^{k}_{s}\cdot x_n =0$, $\forall k,n\neq 0$ and $\forall s,t$,
all being null vectors pointing in the same light like direction.
Thus the wave function should have the form
$e^{il_0\cdot x_0}\prod\Psi_k$. The last term in (\ref{tr}) requires
$\Psi^{\prime\prime}_K (\chi_K) =0$. This is satisfied if $\Psi_K$ is
given by the massless solution (\ref{massl}) with $D_K=0$. Notice 
that eq. (\ref{tr}) is also
satisfied  if we choose $x_K^2=0$ instead of $\Psi_K =C_K$. Requiring
 the funcion $\Phi$ to satisfy the constraints (\ref{g1}) for all the
other values of $m$ as well, we find that $\Psi_k = C_k$, $\forall k\in
{\cal{A}}$,
 ${\cal{A}}$ being a subset of 
$\{ -K,\ldots,-1,1,\ldots,K\}$ and that $x_i
\cdot x_j =0$, $\forall i,j\in  {\cal B}$, where ${\cal B}$ is the
compliment of ${\cal A}$.  The second conditions give in particular that
$\chi_n = 0$, $\forall n \in {\cal B}$. 
So again the only
solution which survives is the {\it massless particle}.
\be
\Phi (x) = \left (\prod_k C_k\right ) e^{il_0\cdot x_0},\nonumber
\ee
with $k\in{\cal A}$.

\begin{flushleft}
\subsection{Case II}
\end{flushleft}

We are now going to solve the equations (\ref{cc1}) and (\ref{c2}). In
the wave function sector (\ref{wf}), the solution may be written as
\be  
|M\rangle &=&|p\rangle ,\nonumber\\
\langle M^{\prime}| &=& \mbox{}_{x}\langle 0|Y(p),\nonumber
\ee  
and the inner product gives
\be
\langle phys^{\prime},0|phys,0\rangle = Y(p).\nonumber
\ee  
Equations (\ref{cc1}) and (\ref{c2}) then imply the relations
\be
\sum_{k=-\infty}^{+\infty}{p_{k}}_{\mu}{p_{m-k}}^{\mu}Y(p)&=&
0,\label{p1}\\
\sum_{k=-\infty}^{+\infty}k{p_{m-k}}^{\mu}
\frac{\partial Y(p)}{\partial {p_{-k}}^{\mu}}&=&0\label{p2}.
\ee

We take  a finite number of modes $s$ again.  For $m=2s$, equation
(\ref{p1}) gives
\be  
&&\sum_{k=-s}^{s}p_{k}\cdot p_{2s-k}Y(p) =p^{2}_{s}Y(p)
=0\nonumber\\ 
&\Rightarrow &p^{2}_{s}=0.\nonumber
\ee  
For $m=2s-1$ we find that $p_{i}\cdot p_{j}=0$, $\forall
i,j\in[s,s-1]$. We can work in the same way to prove by induction that
\be  
p_{i}\cdot p_{j}=0,\qquad \forall i,j\in[-s,s]\nonumber
\ee  
all pointing in the same light-like direction. Therefore, the
function $Y(p)$ should have the form
\be  
Y(p) = Y(a_{s}\cdot p_{-s},a_{s-1}\cdot p_{-s+1},\ldots,a_{-s}\cdot
p_{s}).\nonumber
\ee  
The spectrum is  {\it massless} again since from (\ref{mass}) we
deduce that
$M^{2}Y(p)=0$.

A physical state should also satisfy the relations (\ref{p2}). For
$m=2s$ we get
\be  
&&\sum_{k=-s}^{+s}k{p_{2s-k}}^{\mu}
\frac{\partial Y(p)}{\partial {p_{-k}}^{\mu}}=0\Rightarrow
s{p_{s}}^{\mu}
\frac{\partial Y(p)}{\partial {p_{-s}}^{\mu}}=0\nonumber\\ 
&\Rightarrow
&p_s\cdot a_s =0\Rightarrow p_{-s}\cdot a_s =0.\nonumber
\ee  
Working in the same way as before we can prove that $a_{k}\cdot
p_{-k}=0$,
$\forall k\neq 0$. So we get
\be  
Y(p)= f(a_0\cdot p_0),\nonumber
\ee  
which is the equation of a {\it massless particle}. Thus we find
once again that the string collapses into a massless particle as  in
the classical case. 

Since our theory does not have any parameter to
start with,  it can be argued that the spectrum should 
consist  of massless states only or that it should be continuous. From
the point of view of the limit
$\alpha^{\prime}=1/(\sqrt{2\pi T})\to 
\infty$ of usual string theories, it is clear that all the states in
the leading Regge  trajectory collapse to zero mass. The above
reasoning shows that all daughter trajectories should also collapse to
zero mass.
 
One might ask  how the spectrum would look like if supersymmetry was
also included. In the case of world-sheet supersymmetry it has been
shown in \cite{ps} that the critical dimension is negative and hence
the theory is anomalous. In the case of  space-time supersymmetry on
the other hand one cannot apply the same formalism because of
difficulties that have to do with the  covariant quantization of the
superparticle. 

\bigskip
\begin{flushleft} {\bf Acknowledgments:}I would like to  thank Ulf
Lindstr\"om 
 for  useful comments and  fruitful discussions.
\bigskip
\end{flushleft}

\eject

\end {document}